\documentclass[12pt]{article}
\pdfoutput=1

\usepackage{amsmath,amssymb,amsfonts,epsfig,cite,setspace,bigstrut,framed,eufrak}
\usepackage[all]{xy}
\usepackage{color}
\usepackage{pifont}

\makeatletter \@addtoreset{equation}{section} \makeatother
\renewcommand{\theequation}{\thesection.\arabic{equation}}

\begin{document}

\vskip 0.25in

\newcommand{\todo}[1]{{\bf\color{blue} !! #1 !!}\marginpar{\color{blue}$\Longleftarrow$}}
\newcommand{\nn}{\nonumber}
\newcommand{\comment}[1]{}
\newcommand\T{\rule{0pt}{2.6ex}}
\newcommand\B{\rule[-1.2ex]{0pt}{0pt}}

\newcommand{\CO}{{\cal O}}
\newcommand{\cI}{{\cal I}}
\newcommand{\cM}{{\cal M}}
\newcommand{\cW}{{\cal W}}
\newcommand{\cN}{{\cal N}}
\newcommand{\cR}{{\cal R}}
\newcommand{\cH}{{\cal H}}
\newcommand{\cK}{{\cal K}}
\newcommand{\cT}{{\cal T}}
\newcommand{\cZ}{{\cal Z}}
\newcommand{\cO}{{\cal O}}
\newcommand{\cQ}{{\cal Q}}
\newcommand{\cB}{{\cal B}}
\newcommand{\cC}{{\cal C}}
\newcommand{\cD}{{\cal D}}
\newcommand{\cE}{{\cal E}}
\newcommand{\cF}{{\cal F}}
\newcommand{\cA}{{\cal A}}
\newcommand{\cX}{{\cal X}}
\newcommand{\IA}{\mathbb{A}}
\newcommand{\IP}{\mathbb{P}}
\newcommand{\IQ}{\mathbb{Q}}
\newcommand{\IH}{\mathbb{H}}
\newcommand{\IR}{\mathbb{R}}
\newcommand{\IC}{\mathbb{C}}
\newcommand{\IF}{\mathbb{F}}
\newcommand{\IV}{\mathbb{V}}
\newcommand{\II}{\mathbb{I}}
\newcommand{\IZ}{\mathbb{Z}}
\newcommand{\re}{{\rm Re}}
\newcommand{\im}{{\rm Im}}
\newcommand{\tr}{\mathop{\rm Tr}}
\newcommand{\ch}{{\rm ch}}
\newcommand{\rk}{{\rm rk}}
\newcommand{\ext}{{\rm Ext}}
\newcommand{\bi}{\begin{itemize}}
\newcommand{\ei}{\end{itemize}}
\newcommand{\beq}{\begin{equation}}
\newcommand{\eeq}{\end{equation}}
\newcommand{\bea}{\begin{eqnarray}}
\newcommand{\eea}{\end{eqnarray}}
\newcommand{\ba}{\begin{array}}
\newcommand{\ea}{\end{array}}

\newcommand{\CN}{{\cal N}}
\newcommand{\y}{{\mathbf y}}
\newcommand{\z}{{\mathbf z}}
\newcommand{\C}{\mathbb C}\newcommand{\R}{\mathbb R}
\newcommand{\CA}{\mathbb A}
\newcommand{\CP}{\mathbb P}
\newcommand{\cP}{\mathcal P}
\newcommand{\tmat}[1]{{\tiny \left(\begin{matrix} #1 \end{matrix}\right)}}
\newcommand{\mat}[1]{\left(\begin{matrix} #1 \end{matrix}\right)}
\newcommand{\diff}[2]{\frac{\partial #1}{\partial #2}}
\newcommand{\gen}[1]{\langle #1 \rangle}

\newtheorem{theorem}{\bf THEOREM}
\newtheorem{proposition}{\bf PROPOSITION}
\newtheorem{observation}{\bf OBSERVATION}

\def\theequation{\thesection.\arabic{equation}}
\newcommand{\setall}{
	\setcounter{equation}{0}
}
\renewcommand{\thefootnote}{\fnsymbol{footnote}}

\begin{titlepage}
\vfill
\begin{flushright}
{\tt\normalsize KIAS-P17029}\\

\end{flushright}
\vfill

\begin{center}
{\Large\bf Twisted Partition Functions and $H$-Saddles}

\vskip 1.5cm

Chiung Hwang\footnote{\tt chwang@kias.re.kr} and
Piljin Yi\footnote{\tt piljin@kias.re.kr}
\vskip 5mm

{\it School of Physics,
Korea Institute for Advanced Study,  Seoul 02455, Korea}

\end{center}
\vfill

\begin{abstract}
While studying supersymmetric $G$-gauge theories,
one often observes that a zero-radius limit of the twisted
partition function $\Omega^G$ is computed by the partition
function ${\cal Z}^G$ in one less dimensions. We show how this
type of identification fails generically due to integrations
over Wilson lines. Tracing the problem, physically, to saddles
with reduced effective theories, we relate $\Omega^G$ to a sum of
distinct ${\cal Z}^H$'s and classify the latter, dubbed $H$-saddles.
This explains why, in the context of pure Yang-Mills quantum
mechanics, earlier estimates of the matrix integrals ${\cal Z}^{G}$
had failed to capture the recently constructed bulk index
${\cal I}^G_{\rm bulk}$. The purported agreement
between 4d and 5d instanton partition functions, despite
such subtleties also present in the ADHM data, is explained.

\end{abstract}

\vfill
\end{titlepage}

\tableofcontents
\renewcommand{\thefootnote}{\#\arabic{footnote}}
\setcounter{footnote}{0}
\vskip 2cm

\section{Index and Bulk Index}

Perhaps the simplest of the topological observables, available
for supersymmetric theories, is the twisted partition function,
\begin{equation}
\Omega(\beta;z)\equiv {\rm Tr}\,\left[(-1)^{\cal F} \,e^{z F}\, e^{-\beta {\cal H}}\right]
\end{equation}
where the trace is taken over the physical Hilbert space
and $e^{z F}$ denotes, collectively, all admissible chemical
potential terms. Although we will mostly display results with the chemical
potential turned off in this note, these chemical potentials
are implicitly assumed and often indispensable part of
the computation.

A naive expectation about this quantity is the $\beta$-independence,
which would allow evaluation of $\Omega(\beta;z)$ in the
$\beta\rightarrow 0$ limit. This is supported by
the familiar one-to-one mapping between bosonic and
fermionic states, which at least naively follows
from  the superalgebra
\bea
{\cal H}={\cal Q}^2 \ ,\qquad \{(-1)^{\cal F},{\cal Q}\}=0 \ ,\qquad [e^{z F},{\cal Q}] =0 \ .
\eea
The same reasoning would imply that the twisted partition
functions count supersymmetric objects, and hence are
inherently integral. As such, the twisted partition function
would compute the Witten index \cite{Witten:1982df}, or its various refined
generalizations.

As with any powerful and sweeping argument, however,
this comes with caveats.  One finds that $\beta$-dependence
can actually survive unless the Hilbert space is completely
discrete. Instead, the $\beta\rightarrow 0$  limit produces an object
called the bulk index,
\begin{equation}\label{bulk}
{\cal I}_{\rm bulk}(z)\equiv \lim_{\beta\rightarrow 0} \Omega(\beta;z) \ ,
\end{equation}
which can be sometimes an interesting physical object
by itself. If we are aiming at the (refined) Witten
index, a more appropriate limit is
\bea
{\cal I}(z)= \lim_{\beta\rightarrow \infty}\Omega(\beta;z) \ .
\eea
The difference between ${\cal I}$ and ${\cal I}_{\rm bulk}$,
denoted by $\delta {\cal I}$, may be in some cases
computed separately
and combined to give the integral Witten index,
\bea
{\cal I}(z)= {\cal I}_{\rm bulk}(z)+\delta{\cal I}(z) \ .
\eea
Unlike the bulk index, the continuum contribution
$\delta{\cal I}$ has no convenient and universal
computational tools. For pure Yang-Mills quantum
mechanics and also for $\cN=4$ quiver quantum mechanics,
nevertheless, a general pattern has been uncovered \cite{Lee:2016dbm}
and both ${\cal I}_{\rm bulk}$ and $\delta{\cal I}$
for wide classes of theories have been computed \cite{Lee:2016dbm,Lee:2017lfw}.

On the other hand, the usual localization procedure,
which seemingly computes the twisted partition function
$\Omega(\beta;z)$ at some finite and arbitrary $\beta$,
usually computes ${\cal I}_{\rm bulk}(z)$ \cite{Lee:2016dbm}.
A good hint of this is that the resulting $\Omega$ has no
$\beta$-dependence, regardless of specifics of the
theory. As we noted above, the $\beta$-dependence
does in general persist for theories with continuum sectors,
so the localization must be, in secret, computing
a limit of $\Omega(\beta;z)$. The only two logical possibilities
are either $\beta\rightarrow 0$ or $\beta\rightarrow \infty$.
However, given that the final expressions
are integrals of some local functions, $\beta\rightarrow\infty$
is hardly possible, hence we can anticipate
\bea
 {\cal I}_{\rm bulk}(z)\; =\; \Omega(z)\; \equiv \;\Omega\,\bigr\vert_{\rm localization}\ .
\eea
We will later give a more explicit argument supporting
this for gauged quantum mechanics.

Since $\beta$ can be thought of as the Euclidean time
interval, a dimensional reduction to one less dimension
is natural. Indeed, in supersymmetric quantum mechanics recast
of index theorems, for instance as in Alvarez-Gaume's 1d
path-integral derivation of Euler index \cite{AlvarezGaume:1983at},
the contributing
saddle localizes to constant configurations, and the 1d
path integral reduces to ordinary integral over the target
manifold.
Something like this also happens with supersymmetric
gauged quantum mechanics, where $\beta\rightarrow 0$ limit
reduces the twisted partition function to ordinary
integrals over Lie Algebra and matter representations
thereof, which we collectively call the matrix integral.

The lore is, as such, that one can take an additional
scaling limit of the chemical potential $z=\beta z'$
with vanishing $\beta$ and finite $z'$  and find
\begin{equation}
\Omega^G(z)\quad\rightarrow
\quad {\cal Z}^G (z') \ .
\end{equation}
The right hand side means the matrix integral with
the exponent of the measure given by the dimensional
reduction to 0d of the Euclidean action of the 1d theory.
Note that, here, $z=\beta z'$ limit is taken after
the usual $\beta\rightarrow 0$ limit of $\Omega(\beta;z)$
was taken as in (\ref{bulk}).

However, this natural expectation proves to be false for
general gauge theories, and in particular for 1d gauged
quantum mechanics.  When one compactifies a gauge theory on
a circle, the Wilson line emerges as natural low
energy degrees of freedom, associated with the Cartan
torus. The path integral would involve integrations over
such Wilson line variables, yet it is clear that one
will lose their periodic nature if the $\beta\rightarrow 0$
limit is taken first before performing the integration;
the Cartan torus is replaced by the Cartan subalgebra
of infinite extension. Do we then lose a contributing
sector, say, from somewhere on the opposite side of the Cartan
torus? As we will show, the answer is yes: One generically
loses contributing saddles, or loses poles if a localization
is employed, by taking $\beta\rightarrow 0$ limit casually.

The phenomenon is quite general for twisted partition
functions, regardless of details
of the theory or of the spacetime dimensions, as long as
there is a circle ${\mathbb S}^1$ and the associated
Wilson-line variables to integrate over. Whenever one tries
to relate a twisted partition function on ${\mathbb S}^1
\times {\mathbb M}$ to the partition function of the
dimensionally reduced theory on (compact) $ {\mathbb M}$,
one must worry about such extra saddles. In retrospect,
the same mechanism can be seen to be responsible for
how 2d elliptic genus
generically fails to compute 1d Witten index via an
appropriate limit \cite{HKY}. In this note, however,
we will confine ourselves to 1d/0d examples, and derive
precise relations between the two sides. The same
reasoning and derivation are easily applicable
to higher dimensions, especially when the twisted
partition function is computed by a residue formula
in the space of (complexified) Wilson lines.

\section{Preliminary: An Old Story}

This finding will also resolve an old mystery surrounding
Witten index computations of supersymmetric Yang-Mills
quantum mechanics (SYMQ). These are ${\cal N}=4,8,16$ SYMQ,
respectively obtained from the dimensional reduction of
minimally supersymmetric Yang-Mills theory in $D$-dimensions
with $D=4,6,10$ \cite{Ivanov:1990jn,Danielsson:1996uw}.
Let us start with a review of the argument for
${\cal I}_{\rm bulk}^G \rightarrow {\cal Z}^G $
for this
simplest class of supersymmetric gauged quantum mechanics.
Interestingly enough, this subtlety does not plague
$SU(N)$ cases, namely that of $N$ D0 branes in the type
IIA theory \cite{Witten:1995ex}, for which this identification
was originally derived \cite{Yi:1997eg,Sethi:1997pa}. The content
of this section is borrowed from Ref.~\cite{Yi:1997eg}.

We start with $\cN\ge 4$ pure Yang-Mills quantum mechanics
for arbitrary simple group $G$, whose dimension is denoted
as $g$. $D-1$ bosonic $X_i$ and their canonical conjugates,
in the adjoint of $G$,
obey
\begin{equation}
[\pi^a_i, X_j^b]=-i\delta^{ab}\delta_{ij}\ .
\end{equation}
The spinor consists of ${\cal N}$ adjoint (real)
fermions $\Psi^a_\beta$ obeying
\begin{equation}
\{\Psi^a_\alpha,\Psi^b_\beta\}
=\delta^{ab}\delta_{\alpha\beta} \ ,
\end{equation}
which form $(g{\cal N})$-dimensional Clifford algebra.
The Hamiltonian is
\begin{equation}
{\cal H}=\frac{1}{2}\pi^a_i\pi^a_i-\frac{1}{2}X^a_iK^a_i
+\frac{1}{4}[X_i, X_j]^2 \ ,
\end{equation}
where the sum is over $i,j=1,2,\dots, D-1$ as well
as gauge indices $a=1,2,\dots,g$ and the fermion
bilinears, $K_i^a$, are part of
\begin{equation}
K^a_\mu = if_{abc}\Psi^b\gamma_\mu\Psi^c \ , \qquad \mu=1,\dots,D \ ,
\end{equation}
with appropriate Dirac matrices $\gamma_\mu$.

The ${\cal N}$ supersymmetries are generated by
\begin{equation}
{\cal Q}_\alpha=\gamma_{i\alpha\beta}\Psi^a_\beta\pi^a_i-\frac{1}{2}\gamma_{ij\alpha
\beta}f^{abc}\Psi_\beta^aX_i^bX_j^c \ ,
\end{equation}
and the adjoint thereof, and lead to the twisted
partition function,
\begin{equation}
\Omega^G(\beta)\equiv {\rm Tr}\,\left[(-1)^{\cal F}e^{z F}e^{-\beta {\cal H}}\right]=
\int dX\;\langle X\vert\,{\rm tr}\, (-1)^{\cal F} e^{z F} e^{-\beta {\cal H}} {\cal P}_{G/Z_G}
\,\vert X\rangle \ .
\end{equation}
The projection to the gauge-singlet sector is instituted by
an insertion of
\begin{equation}
{\cal P}_{G/Z_G}=\frac{1}{{\rm vol}(G/Z_G)}\oint_{G/Z_G} d\theta \,e^{i\theta^a G^a}
\end{equation}
with the Gauss constraints
\begin{equation}
G^a=f^{abc}X^b_i\pi^c_i-\frac{i}{2}K^a_D \ .
\end{equation}
We chose to integrate not over $G$ but
over $G/Z_G$, as the center $Z_G$ acts trivially
on the adjoint representation. An important subtlety
related to this projector will be revisited in next section.

With the chemical potential turned off, $z=0$, for
simplicity, the heat kernel expansion suffices,
\begin{equation}
\langle X\vert \,e^{-\beta {\cal H}}\, \vert X'\,\rangle=
\frac{1}{(2\pi\beta)^{g(D-1)/2}}
e^{-(X'-X)^2/2\beta}e^{-\beta(V+H_F)}\left(1+O(\beta)\right)\ ,
\end{equation}
where $V$ is the bosonic potential while $H_F\equiv -X^a_iK^a_i/2$.
The Gauss constraint rotates $\vert X\rangle$ to
$\vert X(\theta) \rangle $, so a small $\beta $ limit of
\begin{equation}\label{1to0}
\frac{1}{{\rm vol}(G/Z_G) (2\pi\beta)^{g(D-1)/2}}
\int dX\oint d\theta\;{\rm tr}_\Psi\, (-1)^{\cal F}
e^{-(X(\theta)-X)^2/2\beta}e^{-\beta\,(V+H_F)}
e^{\theta^a K_D^a/2} \
\end{equation}
needs to be evaluated and thus it is sufficient to
consider regions $X(\theta)\sim X$. An obvious thing
to do is to expand $\theta$ as
\bea \label{expansion}
\theta=\beta \xi \ ,
\eea
whereby one finds
\begin{equation}
\frac{\beta^g}{{\rm vol}(G/Z_G) (2\pi\beta)^{g(D-1)/2}}
\int dX\int[d\xi]\;{\rm tr}_\Psi\, (-1)^{\cal F}
e^{-\beta\, [\xi, X]^2/2-\beta\,V}e^{-\beta\,(H_F-
\xi^a K_D^a/2)} \ .
\end{equation}
Identifying $\xi$ with $X_{D}$, i.e. the Euclideanized $A_0$, we find
that this limit is computed by, with $\beta^{1/4}X\rightarrow X$,
\bea
 \lim_{\tilde\beta\rightarrow 0}
\frac{1}{{\rm vol}(G/Z_G)}\frac{(2\pi)^{g/2}}{(2\pi)^{gD/2}\tilde\beta^{g{\cal N}/2}}
\int dX\;e^{-[X_\mu,X_\nu]^2/4}\;{\rm tr}_\Psi\; (-1)^{\cal F}e^{ \tilde\beta X^a_\mu K^a_\mu/2} \ .
\eea
Keeping only the leading power in small $\tilde\beta\equiv\beta^{3/4}$,
we find a $G$-matrix integral with
${\cal N}=2(D-2)$ supersymmetries,
\bea\label{reduction}
{\cal Z}^G \equiv
\frac{1}{{\rm vol}(G/Z_G)}\frac{(2\pi)^{g/2}}{(2\pi)^{gD/2}}
\int dX\, d\Psi\;e^{-[X_\mu,X_\nu]^2/4 + X^a_\mu K^a_\mu/2} \ .
\eea
This line of reasoning has led to the expectation, with chemical
potentials restored,
\bea\label{reduction2}
{\cal I}_{\rm bulk}^G (z)\quad\rightarrow\quad {\cal Z}^G (z') \ ,
\eea
which has been successfully used for $G=SU(N)$.

For other simple gauge groups, however, we will see that
${\cal Z}^G $ captures only part of ${\cal I}_{\rm bulk}^G $.
Computation of ${\cal Z}^G $ was performed for $G=SU(2)$
by Yi \cite{Yi:1997eg} and also by Sethi and Stern \cite{Sethi:1997pa}
and for $SU(N)$ by Moore, Nekrasov, and Shatashvili \cite{Moore:1998et}.
The latter, in particular, introduced a 0d localization method, to be here-in
referred to as the MNS method, which was then generalized
to arbitrary simple groups by Staudacher \cite{Staudacher:2000gx}
and also by Pestun \cite{Pestun:2002rr}.
On the other hand, a localization method for twisted partition
function of general gauge theories was derived from the first principle
by Hori, Kim, and Yi (HKY) \cite{HKY,JK} and, with this,
${\cal I}_{\rm bulk}^G $ was recently computed \cite{Lee:2016dbm}.
These two sets of answers disagree, except for $SU(N)$'s.

One logical
possibility is that the MNS method with its ad hoc, if
elegant, prescription of the final contour integrals, is
inadequate for general $G$ and fails to compute the matrix
integral correctly. However, we have  confirmed that the
first-principle contour derived along the line of HKY also
give the same set of numbers for ${\cal Z}^G $,
which motivated the study in this note. In the following,
we will show why the above expectation fails,  how
the $SU(N)$ case evaded this subtlety, and also recover
${\cal I}^G_{\rm bulk}$ as a sum of ${\cal Z}^H $'s
with $H$ certain subgroups of $G$.

\section{Localizations and Missing Residues}\label{missing}

It is instructive to recall the gauge projector
\begin{equation}
{\cal P}_{G/Z_G}=\frac{1}{{\rm vol}(G/Z_G)}\oint_{G/Z_G} d\theta \,e^{i\theta^a G^a} \ ,
\end{equation}
which is hardly a unique choice.
Since the center $Z_G$ of $G$ acts trivially on
adjoints, we can equally use
\begin{equation}
{\cal P}_{G}=\frac{1}{{\rm vol}(G)}\oint_{G} d\theta \,e^{i\theta^a G^a} \ .
\end{equation}
Since both project out all unphysical states and keep
all gauge invariant states, there should be no
difference between the two such choices.
How does such an ambiguity manifest in actual
evaluation? As we outlined above, one
canonical way to evaluate relies on the straightforward
$\beta\rightarrow 0 $ limit with an identification
\bea
\label{0}\theta = \beta X_D \ .
\eea
As such, one obtains a $G$-matrix integral such as
(\ref{reduction}), whose integrand is seemingly oblivious to
whether one started with ${\cal P}_G$ or ${\cal P}_{G/Z_G}$.
This cannot be right, however, since these two give an
identical integral but with different overall factors,
$1/{\rm vol}(G)$ or $1/{\rm vol}(G/Z_G)$, respectively.

The resolution to this is  obvious, though.
When one uses ${\cal P}_G$ in place of ${\cal P}_{G/Z_G}$,
the $\beta$ expansion (\ref{0}) is not the only possible
one. Rather one must also consider
\bea\label{G-saddle}
\theta = \Theta_G + \beta X_D
\eea
where $e^{i\Theta_G}$ is an element of $Z_G$. No dynamical
variable is affected by the center $Z_G$ in pure Yang-Mills,
so we have $|Z_G|$-many gauge-equivalent saddle points.
Individual saddles are infinitely separated from each other
in the limit $\beta\rightarrow 0$, yet we must sum them
up since we are computing a limit of 1d quantity rather
than 0d quantity. This sum recovers the extra overall
factor of $|Z_G|$, only to be canceled by the same
factor in ${\rm vol}(G)= |Z_G|\cdot {\rm vol}(G/Z_G)$.

In fact, this ambiguity goes beyond $G$ vs $G/Z_G$.
Sometimes, we find it simpler to parameterize the group
by a multiple cover. A main example of this for us would
be $F_4$, which has no center and whose Cartan torus
has the natural volume $(2\pi)^4/8$. Yet, it is easier
to deal with $F_4$ if we pretend that each Cartan
parameters span $[0,2\pi)$ independently. If we choose
to do this, the integral range in the projector become
8-fold larger and must be compensated by the factor
$1/(8\cdot{\rm vol}(F_4))$.  Therefore, the required list
of $e^{i\Theta_G}$ is determined as much by how one parameterizes
$G$ as by the abstract group structure and the field
content of the theory.

For such reasons, the normalization issue is generally
more subtle with the matrix integral than its 1d counterpart,
if one wishes to recover ${\cal I}_{\rm bulk}^G$ via matrix
integral. Thankfully, this has been worked out in the
past for SYMQ. A particularly powerful version is via a 0d
localization which leaves only rank-many contour integrals
as \cite{Moore:1998et,Staudacher:2000gx,Pestun:2002rr}
\bea\label{0dL}
{\cal Z}^G (z') =\frac{|{\rm det}(Q^G_{ab})|}{|W_G|}\int_{{\cal C}'} \frac{d^{r_G} u'}{(2\pi i)^{r_G} } \; f_G(u';z')
\eea
with the Weyl group $W_G$ and $r_G={\rm rank}(G)$. The generalization of
$|Z_G|$ factor sits in ${\rm det}(Q^G_{ab})$
where $|Q^G_{ab}|$ is the matrix spanned by the
simple roots. Potential rescaling of $u's$ in the
measure is counteracted by this determinant, and,
for example, $|{\rm det}(Q^G_{ab})|$ equals
$|Z_G|$ for classical gauge groups with the
defining representations normalized to have ``unit"
charges.

Let us come back to 1d and discuss how this normalization
factor shows up in the 1d localization procedure. In the
latter the electric coupling constant, $e^2$, instead
of $\beta$, is taken to zero,
\bea
\Omega^G (z)\equiv \lim_{e^2\rightarrow 0}\Omega^G(\beta;z) \ .
\eea
For theories with no other parameters, $e^{2/3}\beta$ is
the only dimensionless parameter, so clearly
the localization process $e^2\rightarrow 0$ has to
compute the $\beta\rightarrow 0$ limit,
\bea
\Omega^G (z) = {\cal I}_{\rm bulk}^G(z) \ .
\eea
Fayet-Iliopoulos constants could have complicated
this identificaton, but it has been observed that
FI constants need to be scaled to infinite, ahead
of any $e^2$ scaling, in order to minimize
continuum contributions from the flat Coulombic
directions \cite{HKY}. The 1d localization assumes
such a limit, hence the above identification remains
valid despite FI constants.
Indeed, a perfect match between $\Omega^G$ as the bulk
index \cite{Lee:2016dbm} and independently computed
defect terms $\delta {\cal I}^G$
\cite{Yi:1997eg,Green:1997tn,Kac:1999av,Lee:2016dbm} has been
established, such that
the combination  ${\cal I}^G=\Omega^G+\delta{\cal I}^G$
is integral as the true Witten index should be.
The resulting Witten indices matched M-theory
predictions \cite{Witten:1995ex,Hanany:1999jy} not only
for $G=SU(N)$ but also for $Sp(N)$, and $O(N)$ \cite{Lee:2017lfw}.

The 1d localization for $\Omega^G={\cal I}_{\rm bulk}^G$
also gives a contour integral \cite{HKY},
\bea\label{1dL}
\Omega^G (z) =\frac{1}{|W_G|}\int_{\cal C} \frac{d^{r_G}  u}{(2\pi i)^{r_G} } \; g_G(e^{u};e^z) \ .
\eea
The contour ${\cal C}$ and the integrand $g_G$
reduces to ${\cal C}'$ and $f_G\cdot \beta^{-r_G} $, respectively,
by a dimensional reduction process with
\bea
u=\beta u' \ , \qquad z=\beta z'
\eea
in the limit of $\beta \rightarrow 0 $ while maintaining
finite $u'$ and $z'$.
This brings (\ref{1dL}) to (\ref{0dL}), with a caveat.
Because of the scaling, 0d localization contour $\cC'$
can keep only a subset of $\cC$, namely only those that
can be shrunken to an infinitesimal neighborhood near
$e^u=1$.  Any other part of $\cC$,
say, with  nonzero phases of $e^{u}$, cannot survive
the limit.\footnote{
Strictly speaking, the original contour prescription
by MNS generally differs from this limit, $\cC'$,
when $\cC$ is the one derived from the first principle
\cite{Benini:2013nda,Benini:2013xpa,HKY}. For examples
shown in this note, however, this difference does not
seem to matter.}

Now we are ready to discuss
how the overall factor $|{\rm det}(Q^G)|$ of
(\ref{0dL}) is hidden in  (\ref{1dL}). The  gauge-variables
$u$ are, unlike $u'$ that live in ${\mathbb C}^r$,
periodic variables living in $({\mathbb C}^*)^r$.
As such, the 1d localization formula secretly assumes
$2\pi i$ periodic $u$-variables, independent of
one another, and ignores possible discrete division,
such as by $Z_G$. Therefore, the poles of $g_G(e^u;e^z)$,
contours around which would constitute ${\cal C}$,
come in $|{\rm det}(Q^G)|$-many multiplets,
separated from one another by $e^{i\Theta_G}$
shifts. In the 1d localization, therefore, one must
sum over these identical residues, and this emulates
the factor $|{\rm det}(Q^G)|$. In the 0d localization,
on the other hand, these poles are located infinitely
far away from each other in $u'$ planes, so ${\cal Z}^G $
will miss these residues, which is then remedied
by the overall numerical factor as in (\ref{0dL}).

Also this comparison neatly resolves the apparent
normalization ambiguity of the measure on the 0d side.
The 1d Hilbert space trace has no such ambiguity and
the simplest convention is to demand gauge charges
normalized so that an individual $u$ variable has $2\pi i $  period.
For classical groups, it is easy to show the numerical
factor $|{\rm det}(Q^G)|$ equals $|Z_G|$, while for $F_4$,
e.g., the same normalization requires the roots to be
of the length-squared 4 and 8, resulting in
${\rm det}(Q^{F_4}_{ab})=8$. The latter number has precisely
the same origin as 8 mentioned earlier, and manifests
in the 1d localization as 8-fold degeneracy of
poles.\footnote{Ref.~\cite{Pestun:2002rr}
used a different $Q^{F_4}$ normalization, which was
nevertheless countered correctly by the measure, $du'$.}

The reduction from  the 1d twisted partition function
to the 0d matrix side is fraught with other dangers,
however; Once we accept the possibility of additional
saddles, shifted by $e^{i\Theta_G}$, we must
also ask whether there might be a different type
of saddles, not gauge-equivalent to the one at origin
but contributing to ${\cal I}^G_{\rm bulk}$.
In view of how $e^{i\Theta_G}$-shifted saddles
correspond to missing residues in 0d, it is easy to
imagine that such new type of saddles, if any, will
also manifest as missing residues when we reduce
(\ref{1dL})  to (\ref{0dL}).

As is evident from the above 0d vs 1d comparison,
some of the poles are lost in the process of $\beta\rightarrow 0$ limit.
Consider a pair of poles for pure $Sp(1)$ theory, related
by a center $Z_{Sp(1)}={\mathbb Z}_2$. If they are
located at $e^{u_*}=e^z,-e^z$, the first of the two
would survive the limit but not the latter, as we
rescale  $z=\beta z'$ and $ u=\beta u'$ with finite
$z'$ and $u'$. This particular loss of poles is
innocuous since, as we saw above, it can be corrected
by a factor 2 associated with $|Z_{Sp(1)}|=2$.
However, given a doublet of poles in the 1d
localization, there is no guarantee that one of
them does survive the 0d limit. Suppose that the pair
happen to sit at $e^{u_*}=ie^{z},-ie^{z}$; both of them
would have been pushed out to infinity as we go over to
the $u',z'$ variables. While this does not actually
happen for pure $Sp(1)$, something similar does happen
generically for $Sp(N\ge2)$ and many other simple groups
of rank two and higher.\footnote{
With fundamental matters present, a similar mismatch of
poles between 0d and 1d can happen even for rank-one theories,
 as we will encounter later in the ADHM examples.}

\begin{table}
\centering
\begin{tabular}{c|cc}
$\quad{\cal N}=4\quad $ & $\qquad {\cal I}^G_{\rm bulk}(0) =\Omega^G(0) \quad$ & $\quad{\cal Z}^{G}$ \\ \\ \hline \\
$SU(N)$ & $\frac{1}{N^2}$ & $\frac{1}{N^2}$ \\ \\
$Sp(2)$ & $\frac{5}{32}$ &  $\frac{9}{64}$ \\ \\
$Sp(3)$ & $\frac{15}{128}$ & $\frac{51}{512}$ \\ \\
$Sp(4)$ & $\frac{195}{2048}$ & $\frac{1275}{16384} $ \\ \\
$Sp(5)$ & $\frac{663}{8192}$ & $\frac{8415}{131072} $ \\ \\
$Sp(6)$ & $\frac{4641}{65536}$ & $ \frac{115005}{2097152}$ \\ \\
$Sp(7)$ & $\frac{16575}{262144}$ & $ \frac{805035}{16777216}$ \\ \\
$SO(7)$ & $\frac{15}{128}$ & $\frac{25}{256}$ \\ \\
$SO(8)$ & $\frac{59}{1024}$ & $\frac{117}{2048}$ \\ \\
$SO(9)$& $\frac{195}{2048}$ & $\frac{613}{8192}$ \\ \\
$SO(10)$ & $\frac{27}{512}$ & $\frac{53}{1024}$ \\ \\
$SO(11)$& $\frac{663}{8192}$ & $\frac{1989}{32768}$ \\ \\
$SO(12)$ & $\frac{1589}{32768}$ & $\frac{6175}{131072}$ \\ \\
$SO(13)$& $\frac{4641}{65536}$ & $\frac{26791}{524288}$ \\ \\
$SO(14)$ & $\frac{1471}{32768}$ & $\frac{5661}{131072}$ \\ \\
$SO(15)$& $\frac{16575}{262144}$ & $\frac{92599}{2097152}$ \\ \\
\hline \\
$G_2$ & $\frac{35}{144}$ &  $\frac{151}{864}$ \\ \\
$F_4$ & $\frac{30145}{165888}$ &  $\frac{493013}{ 3981312}$ \\ \\
\end{tabular}
\caption{${\cal I}^G_{\rm bulk}$ vs. ${\cal Z} ^G$,
with $SO(3)\simeq SU(2)\simeq Sp(1)$,  $SO(4)\simeq SU(2)\times SU(2)$, $SO(5)\simeq Sp(2)$,
and $SO(6)\simeq SU(4)$ understood}
\label{numbers}
\end{table}

This means that 0d localization computation of a $G$-matrix
integral ${\cal Z}^G $ will generally miss residues which
would have contributed to the 1d computation, ${\cal I}^G_{\rm bulk}$,
and the loss cannot be compensated by an overall numerical factor.
Indeed, a recent computation \cite{Lee:2016dbm} of 1d twisted
partition functions for general simple gauge group $G$ gave
answers different from those of the 0d matrix integral,
with the exception of $G=SU(N)$ theories. In Table \ref{numbers},
we list these two sets of numbers for ${\cal N}=4$ SYMQ.
The numbers in the second column are borrowed from Ref.~\cite{Lee:2016dbm}
which worked out the bulk index ${\cal I}_{\rm bulk}^{G}=\Omega^{G}$
and the Witten index ${\cal I}^{G}$, for $\cN=4,8,16$ and general
$G$. Here, we took the unrefined limit, for the comparison with
${\cal Z}^G $. The numbers in the third column are newly
computed using 0d localization of ${\cal Z}^G $ whose
$z'$-dependence drops out for $\cN=4$.

The latter set of numbers
agree with older analytical results by Staudacher \cite{Staudacher:2000gx}
and by Pestun \cite{Pestun:2002rr} as well as with the Monte Carlo estimates by
Krauth and Staudacher \cite{Krauth:2000bv}
within the latter's error bars. Interestingly,
${\cal Z}^G $ is consistently smaller than
${\cal I}_{\rm bulk}^{G}$ whenever the two disagree. In
next section, we will dig deeper and unravel the precise
physical reason behind such disagreements. It is worthwhile
to repeat here that such disagreements are not confined to
SYMQ but prevalent phenomena for twisted partition functions
of gauge theories in various dimensions. We will also discover
the correct identities between ${\cal I}_{\rm bulk}^G$'s and
${\cal Z}^G $'s, in next section, and test them against explicit
0d and 1d computations in Section 5.

\section{$H$-Saddles}
\label{sec:H-saddles}

The statement that a limit of ${\cal I}_{\rm bulk}^G$ is computed by
${\cal Z}^G $, although widely accepted in the
community, must be therefore revised. We will presently find, in general,
\bea
{\cal I}^{G}_{\rm bulk}(z=\beta z')\biggr\vert_{\beta\rightarrow 0}
= {\cal Z}^{G} (z')+\cdots
\eea
where the ellipsis denotes additional contributions due
to saddles which are not gauge-equivalent to the one at
origin. See Eq.~(\ref{correct}) for the precise formula.
For the rest of this note, we will study and
catalog such additional contributions, to be called
$H$-saddles.

To understand when and how such saddles appear, let us step
back to the expression (\ref{1to0}) for the unrefined ${\cal I}^G_{\rm bulk}$,
\begin{equation}\label{one2zero}
\sim \beta^{-g(D-1)/2}
\int dX\oint d\theta\;
e^{-(X(\theta)-X)^2/2\beta-\beta V}{\rm tr}_\Psi\, (-1)^{\cal F}
e^{-\beta H_F+\theta^a K_D^a/2}
\end{equation}
and  the subsequent expansion of $\theta$, (\ref{expansion}),
around $\theta=0$. Could there be other saddles
in this $\beta\rightarrow 0$ limit? We already noted that
gauge-equivalent saddles at, $\theta =\Theta_G +\beta X_D$,
infinitely far away from the 0d perspective, must be summed
over if we had chosen to use ${\cal P}_{G}$ instead of
${\cal P}_{G/Z_G}$. Each of these saddles leads to the
same matrix integral as (\ref{one2zero}), so an overall
numerical factor was invented and effectively took care
of them, instead. Let us write similarly,
\bea\label{Theta}
\theta =\Theta +\beta X_D
\eea
and ask for what other $\Theta$'s can there be a contribution
to ${\cal I}^G_{\rm bulk}$.

\subsection{The 0d Limit and $H$-Saddles}\label{Hsaddle}

Turning on $\Theta$ is analogous to a Wilson line symmetry
breaking, so we split the Lie Algebra ${\mathfrak g}$ of $G$ into
${\mathfrak h}$ of the unbroken subgroup $H\subset G$ and the
rest ${\mathfrak j}$. The commutators obey
\bea
[{\mathfrak h},{\mathfrak h}]\sim {\mathfrak h} \ ,\qquad [{\mathfrak h},{\mathfrak j}]\sim {\mathfrak j}
 \ ,\qquad [{\mathfrak j},{\mathfrak j}]\sim {\mathfrak h}+{\mathfrak j} \ .
\eea
Then, we split the variables into two parts
as
\bea
X_\mu &= & Z_\mu^{\mathfrak h}+Y_\mu^{\mathfrak j} \ , \cr\cr
\Psi &=& \Phi^{\mathfrak h} + \Lambda^{\mathfrak j}\ .
\eea
The superscripts ${\mathfrak h}$ and ${\mathfrak j}$ will be
henceforth suppressed.

The fermion part of the exponent can be schematically written as
\bea
-\beta H_F+\theta^a K_D^a/2 &\sim & \Theta \Lambda\Lambda +\beta Z\Phi\Phi +\beta Z\Lambda\Lambda + \beta Y \Phi\Lambda \ ,
\eea
which we need to bring down $g{\cal N}/2$ times to saturate
the fermionic trace.
Two immediate facts follow from this rough form.
First, since all of $\Lambda$'s couple to $\Theta$, terms
of type $\beta Z\Lambda\Lambda$ are irrelevant for
$\beta \rightarrow 0$ limit,
\bea
-\beta H_F+\theta^a K_D^a/2 &\sim & \Theta \Lambda\Lambda +\beta Y\Lambda\Phi + \beta Z\Phi\Phi \ .
\eea
Second, if $H$ contains a $U(1)$ factor, not all
of $\Phi$ can appear in $Z\Phi\Phi$. Then, for each such decoupled $\Phi$,
the fermionic trace has to bring down one factor of $\beta Y\Lambda\Phi$.
This means, in turn, the $\Lambda$ trace will cost an extra power of
$\beta Y$ than otherwise. Combined with the $\beta$ power counting
in the subsequent $Y$ and $Z$ integration, we find that $\beta\rightarrow 0$
will kill this expansion, regardless of the detail.

Therefore, a saddle contribution around (\ref{Theta}) may
contribute only if the unbroken $H$ is either a simple group
or a product of simple
groups. Such $\Theta$ is possible only at discrete points,
and we must in general sum up
such $H$-saddles if we wish to express ${\cal I}_{\rm bulk}^G$ as matrix integrals.
It remains to show, though, that such an ``$H$-saddle"
with nonvanishing fermionic trace does contribute
to ${\cal I}_{\rm bulk}^G$.
Let us first see how the massive bosonic degrees of
freedom, associated with the broken part ${\mathfrak j}$, contribute. The fermionic
trace over $\Lambda$ produces no factors of $\beta$ or $X$
in the leading terms, and leaves only $\Phi$ trace.
Since we can consider $Y$ and $\Lambda$ ``fast" variables, its
integration will lead to an additive contribution to ${\cal I}_{\rm bulk}^G$
so that
\bea\label{G2H}
{\cal I}_{\rm bulk}^G\quad\rightarrow\quad
\sum_{H \subset G} \int dZ\, d\Phi\;{\cal O}_{G;H}(Z)\, e^{-[Z,Z]^2/4 +Z_\mu K_\mu(\Phi)/2 }\ ,
\eea
for some operator ${\cal O}_{G;H}(Z)$ of fixed degree.
We will collect the power of $\beta$ by starting with
(\ref{one2zero}) and show that such $H$-saddle contributions
generically survive $\beta\rightarrow 0$ limit.

The explicit factor of $\beta$ in front of (\ref{one2zero})
can be conveniently split into,
with $h\equiv {\rm dim}H$,
\bea\label{power}
\frac{\beta^h}{\beta^{h(D-1)/2}}\cdot \frac{1}{\beta^{(g-h)(D-1)/2}}\cdot \beta^{g-h} \ .
\eea
We already saw that the fermionic part contributes power of $\beta$
only via $\Phi$ trace, which together with the first factor
above cancels out in the transition to an $H$-matrix integral. It remains
to count the powers of $\beta$ generated by $Y$ integration.
The bosonic part of the exponent can be schematically grouped
as follows,
\bea
\frac{1}{\beta}(X(\theta)_i-X_i)^2&\rightarrow & \frac{1}{\beta}\left(\Delta_\Theta Y_i
+ \beta[Y_D,Y_i]+\beta [Y_D, Z_i] + \beta [Z_D, Y_i] +\beta [Z_D,Z_i]\right)^2 \ ,\cr\cr
\beta[X_i,X_j]^2&\rightarrow & \beta\left([Y_i,Y_j]+[Y_i,Z_j]+[Z_i,Y_j]+[Z_i,Z_j]\right)^2 \ .
\eea
Since $\Delta_\Theta Y_i\equiv Y(\Theta)_i-Y_i$ is of order $\beta^0$, we can
drop some of higher order $Y_i$ terms, leaving behind,
\bea
\frac{1}{\beta}(X(\theta)_i-X_i)^2&\rightarrow & \frac{1}{\beta}\left(\Delta_\Theta Y_i
+\beta [Y_D, Z_i]\right)^2 + \beta  [Z_D,Z_i]^2 \ ,\cr\cr
\beta[X_i,X_j]^2&\rightarrow & \beta\left([Y_i,Z_j]+[Z_i,Y_j]\right)^2+\beta [Z_i,Z_j]^2 \ .
\eea
Terms involving $Z_\mu$ variable are needed to constitute $H$-matrix
integral, so we only need to consider terms with $Y$, and the
integration thereof.

$Y_i$ integration generates $\beta^{(g-h)(D-1)/2}$
which cancels the second factor of (\ref{power}), and also
replaces $Y_i$ inside $[Y_i,Z_j]$ by $\beta\Delta_\Theta^{-1}
[Z_i, Y_D]$. If the action of $\Delta_\Theta$ on $Y_i$
is diagonal, we can further organize
\bea
\beta \left([Y_i,Z_j]+[Z_i,Y_j]\right)^2 &
\sim & \beta^3 \left( [[Z_i,Z_j],Y_D]\right)^2 \ .
\eea
The subsequent integration of $Y_D$ generates prefactors,
which, combined with the third factor in (\ref{power}),
produce
\bea
\sim \frac{1}{\beta^{(g-h)/2}}\cdot \frac{1}{Z^{2(g-h)}} \ .
\eea
With the standard rescaling $\beta^{1/4}Z\rightarrow Z$,
we see finally that the extra power of $\beta$ cancels out,
and
\bea
{\cal O}_{G;H}(Z)\sim \frac{\beta^0}{Z^{2(g-h)}} \ .
\eea
The resulting $H$-matrix integral has no reason
to vanish whatsoever, and thus must contribute to
${\cal I}^G_{\rm bulk}$ additively.
We conclude that, in general, ${\cal Z}^G $
cannot by itself compute  the unrefined limit of ${\cal I}^G_{\rm bulk}$.

\subsection{Recovering $\Omega^G(z)$ from $H$-Saddles }

While we have demonstrated how $H$-saddles can contribute
additively to $\Omega^G={\cal I}^G_{\rm bulk}$, their evaluation
is another matter. Such $H$-saddles must account for the
difference, e.g., for $1/64$ for $\cN=4$ $Sp(2)$,
\bea\label{Sp2}
\Omega^{Sp(2)} \biggr\vert_{\rm unrefined} = {\cal Z}^{Sp(2)} +\frac{1}{64} \ .
\eea
Can we account for such differences precisely by
evaluating $H$-saddle contributions, saddle by saddle?
On the other hand, we already noted how the such additive
difference manifests as the missing residue phenomena between the 1d
localization and the 0d localization. It is thus natural to ask
if one can establish a precise relation between these
two and thereby compute individual $H$-saddles via localization.

For gauged quantum mechanics with at least two
supersymmetries, HKY derived the residue formula
for $\Omega$,
\bea\label{jk-formula}
\Omega
=\frac{1}{|W_G|}\;{\text{JK-Res}}_\eta\;\frac{g(t;{\bf y},\cdots)}{\prod_s t_s}\;{\rm d}^r t \ ,
\eea
where $(t_1,\dots,t_r)$ parameterize the  Cartan torus,
$({\mathbb C}^*)^{r}$. For $\cN\ge 4$, with
$SU(2)\times U(1)$ $R$-symmetry, the functional
determinant $g$ takes the form,
\bea\label{det}
g(t;{\bf y}, \cdots)
&=&\left(\frac{1}{{\bf y}-{\bf y}^{-1}}\right)^{r}
\prod_{\alpha}\frac{t^{-{\alpha/ 2}}-t^{\alpha/ 2}}{
t^{\alpha/ 2}{\bf y}^{-1}-t^{-{\alpha/ 2}}{\bf y}} \cr\cr\cr
&&\times \prod_{i}\frac{t^{-Q_i/2}x^{-{F_i/ 2}}{\bf y}^{-\left({R_i/ 2}-1\right)}
-t^{Q_i/2}x^{{F_i/ 2}}{\bf y}^{{R_i/ 2}-1}}{
t^{Q_i/2}x^{{F_i/ 2}}{\bf y}^{R_i/ 2}
-t^{-Q_i/2}x^{-{F_i/ 2}}{\bf y}^{-{R_i/ 2}}}
\label{4g}
\eea
where $\alpha$ runs over the roots of the gauge group
and $i$ labels the individual chiral multiplets, with
$U(1)$  $R$-charge $R_i$, the gauge charge $Q_i$ under
the Cartan. The chemical potential terms
asscoiated with Cartan of the flavor group are
denoted collectively as $x^{F_i}$. For detailed
derivation and description of this JK residue \cite{JK} formula
as well as for how to select and use the auxiliary
parameters $\eta$, please see the section 4 of Ref.~\cite{HKY}.

This arises from the $e^2\rightarrow 0$ limit of the
path integral version of $\Omega(\beta;z)$,
\bea
\Omega(\beta;z) =\int [d \cA_0\,d \cX_i\, \cdots]\;
\exp\left({-\int_0^\beta {\cal L}_{\rm Euclidean}(\cA_0,\cX_i,\cdots; {\bf z}\equiv 2\log{\bf y},\cdots)}\right) \
\eea
where, as we already noted, the
$\beta$-dependence is implicitly removed by this
localization process. On the other hand, the naive
$\beta\rightarrow 0$ limit of this path integral is
\bea
{\cal Z} =\int dX_D\,dX_i\, \cdots\;
\exp\left({-{\cal L}_{\rm Euclidean}(X_D,X_i,\cdots; {\bf z}', \cdots )}\right)
\eea
obtained by restricting the fields to the constant
configurations
\bea
\cA_0 \rightarrow X_D \ , \quad \cX_i\rightarrow X_i \ ,\quad \cdots
\eea
and expanding the chemical potentials as
\bea
\log{\bf y} =\beta {\bf z}'/2
\eea
with ${\bf z}'$ kept finite, and similarly for flavor chemical
potentials $x$.

Evaluating the latter matrix integral, one obtains
the 0d contour integral formula referred to in the previous
section, as follows: With the 1d localization formula, take
$\beta\rightarrow 0$ limit on the integrand first,
\bea\label{naive}
&&\frac{1}{|W_G|}\;{\text{JK-Res}}_\eta\; \lim_{\beta\rightarrow 0}
\beta^r g_G(t=e^{\beta u'};{\bf y}=e^{\beta {\bf z}'/2},\cdots)\;{\rm d}^r u'\cr\cr\cr
&\rightarrow &\frac{1}{|W_G|}\;{\text{JK-Res}}_\eta'\; f_G(u';{\bf z}',\cdots) \;{\rm d}^r u'\ ,
\eea
where we took care to put a prime in the latter JK-Res
to emphasize that not all available poles of $g$
survive this limit. This misses the other gauge-equivalent
saddles, (\ref{G-saddle}), so more generally
we must also include  contributions from
\bea
\cA_0 \rightarrow \frac{\Theta_G}{\beta}+ X_D \ , \quad \cX_i\rightarrow X_i \ ,\quad \cdots
\eea
for the Wilson lines $e^{i\Theta_G}$. We sum over
these and find
\bea\label{better}
{\cal Z}^G  &=& \sum_{\Theta_G}
\frac{1}{|W_G|}\;{\text{JK-Res}}_\eta\;
\lim_{\beta\rightarrow 0}\beta^r g_G(t=e^{i\Theta_G}
e^{\beta u'};{\bf y}=e^{\beta {\bf z}'/2},\cdots)\;{\rm d}^r u' \cr\cr\cr
&=& \frac{|{\rm det}(Q^G)|}{|W_G|}\;{\text{JK-Res}}_\eta'\;
f_G (u';{\bf z}',\cdots) \;{\rm d}^r u' \ ,
\eea
since $g_G$ is invariant under such shifts. This is
the 0d formula (\ref{0dL}) of the previous section.

It is quite clear that nontrivial $H$-saddles are no
different than $\Theta_G$ saddles, in that they are
merely different kinds of Wilson lines,
\bea
\cA_0 \rightarrow \frac{\Theta}{\beta}+ X_D \ , \quad \cX_i\rightarrow X_i \ ,\quad \cdots
\eea
around which a reduced $H$ theory resides. Taking these
into account as well, one finds
\bea
&&\Omega^G(\beta z')\biggr\vert_{\beta\rightarrow 0}\\ \cr
&=&{\cal Z}^G  + \sum_{\Theta}\frac{1}{|W_G|}\;{\text{JK-Res}}'_\eta\;
\lim_{\beta\rightarrow 0}\beta^r g_G(t=e^{i\Theta}e^{\beta u'};{\bf y}
=e^{\beta {\bf z}'/2},\cdots)\;{\rm d}^r u' \ . \nonumber
\eea
If a pole for $\Omega^G$, say, at $t_*=h({\bf y},x)$,
survives the 0d limit, the pole at $t_*=e^{-i\Theta}h({\bf y},x)$
would be missed by ${\cal Z}^G $ but could be a
contributing pole in the $\Theta$ summand. This way, the latter
sum compute $H$-saddles individually via a 0d localization.

One can go further, in fact. The poles
missed by (\ref{better}) are all such that the argument
of Sinh functions in the denominator of (\ref{det})
are either intact or shifted by some finite angle,
due to
\bea \label{minus}
e^{i\Theta} E_{\alpha} e^{-i\Theta} =e^{i\phi_\alpha} E_{\alpha}
\eea
for some $\phi_\alpha \in(0,2\pi)$.
For pure gauge theories, then, the contributing determinant
factors in $g_G$ fall in two distinct categories. For roots
belonging to the unbroken group $H$, we merely take the 0d
scaling limit as
\bea
\frac{t^{-{\alpha/ 2}}-t^{\alpha/ 2}}{
t^{\alpha/ 2}{\bf y}^{-1}-t^{-{\alpha/ 2}}{\bf y}}
\quad\Rightarrow\quad
\frac{e^{-\beta u'\cdot \alpha/ 2}-e^{\beta u'\cdot \alpha/ 2}}{
e^{\beta (u'\cdot \alpha-{\bf z'})/ 2}-e^{-\beta (u'\cdot \alpha-{\bf z'})/ 2}}
\quad\rightarrow\quad
-\frac{\alpha\cdot u' }{ \alpha\cdot u' -{\bf z}'} \ .
\eea
The broken ones suffer a common and nonzero shift of
the phase both in the numerator and in the denominator,
and, thanks to this, reduces to $-1$ universally in the
0d scaling limit,
\bea
\frac{t^{-{\alpha/ 2}}-t^{\alpha/ 2}}{
t^{\alpha/ 2}{\bf y}^{-1}-t^{-{\alpha/ 2}}{\bf y}}
\quad\Rightarrow\quad
\frac{e^{-\beta u'\cdot \alpha/ 2} e^{-i\phi_\alpha/2}
- e^{\beta u'\cdot \alpha/ 2} e^{i\phi_\alpha/2}}{
e^{\beta (u'\cdot \alpha-{\bf z'})/ 2} e^{i\phi_\alpha/2}
-e^{-\beta (u'\cdot \alpha-{\bf z'})/ 2} e^{-i\phi_\alpha/2}}
\quad\rightarrow\quad -1 \ .
\eea
For contributions from the
adjoint chirals, the same happens, producing $-1$'s
for the latter class in particular. Therefore,
the integrand reduces, at such shifted saddles, to
\bea
\lim_{\beta\rightarrow 0}\beta^r
g_G(t=e^{i\Theta}e^{\beta u'};{\bf y}=e^{\beta {\bf z}'/2},\cdots)
=
\lim_{\beta\rightarrow 0}\beta^r
g_H(t=e^{\beta u'};{\bf y}=e^{\beta {\bf z}'/2},\cdots) \ .
\eea
The saddle contribution at $e^{i\Theta}$ is therefore
nothing but the canonical $H$-matrix integral, except
that the overall group theory factor in front is that
of $G$ rather than that of $H$, which, amazingly, must
be the sole effect of the complicated operator
insertion $\cO(G;H)$ in (\ref{G2H}).

After careful account of $H$-saddles and their Weyl copies,
we arrive at the following universal formula for $\cN=4,8,16$ SYMQ,
\bea\label{correct}
\Omega^G(\beta z')\biggr\vert_{\beta\rightarrow 0} \quad=\quad
{\cal Z}^G (z') + \sum_{H}d_{G:H}\,
\frac{|{\rm det}(Q^G)|/|W_G|}{ |{\rm det}(Q^H)|/|W_H| }\;{\cal Z}^H (z') \ .
\eea
The integer $d_{G:H}$ counts the number of
Wilson lines that leave $H$ unbroken;
it counts Weyl copies of $e^{i\Theta}$
as distinct, modulo $e^{i\Theta_G}$ shift by
left multiplication. Note that, since the whole formula
started with the $G$ theory, the normalization of the
charge matrix $Q^H$ of simple roots for $H$ must be
the one inherited from the $G$ root system.
For classical group $G$, this further simplifies to
\bea\label{correctC}
\Omega^G(\beta z')\biggr\vert_{\beta\rightarrow 0} \quad=\quad
{\cal Z}^G (z') + \sum_{H}\frac{d_{G:H}}{2}\,
\frac{|W_H|}{|W_G| }\;{\cal Z}^H (z') \ .
\eea
Again, the integer $d_{G:H}$ counts the number of
Wilson lines, up to the action of $Z_G$ by left
multiplication. We will show that such a $H$-saddle
sum (and its analog) happen to be absent for $SU(N)$
SYMQ and for $U(k)$ ADHM, but otherwise generically
present for gauged quantum mechanics.

While we concentrated on 1d theories in this note, it is
pretty clear that the phenomena of the missing residues are
prevalent whenever we consider gauge theory on a vanishing
circle, i.e., when we compute the twisted partition function
of a supersymmetric gauge theory on ${\mathbb S}^1\times {\mathbb M}$
and try to relate its limit to partition functions on ${\mathbb M}$.
In fact,  the derivation here is easily extendible, regardless
of the details of the theory or even of the spatial dimension,
dim(${\mathbb M}$), as long as a residue formula involving
the ${\mathbb S}^1 $ Wilson line variables is available.

\section{Classifying $H$-Saddles}

For Yang-Mills theories with adjoint representations
only, a Wilson line, $e^{i\Theta}\in G$, gives a contributing
saddle if and only if it preserves $H$ a product of simple subgroups
of $G$.
One convenient parametrization of the Wilson line is
\bea
\Theta  = 2\pi \sum_s \frac{2k_s^{(\Theta)}}{|\beta_s|^2} \;\vec\mu_s\cdot \vec {\mathbb H}
\eea
with Cartan generators $\vec {\mathbb H}$, simple roots $\vec \beta_s$ and the associated
fundamental weights $\vec\mu_s$. A general positive root
$\vec\alpha_{\{n\}}= \sum_s n_s\vec\beta_s$
of $G$ is in the $H$ root system if and only if
\bea\label{H}
\sum_s k^{(\Theta)}_s n_s \;=\; 0 \;\; {\rm mod}\;\; {\mathbb Z} \ .
\eea
For generic values for $k^{(\Theta)}_s$'s,
it is clear that $U(1)$ generated by $\Theta $ itself will
be a free $U(1)$ in $H$. Only at discrete choices
of $\Theta$, we expect to find contributing saddles.

\subsection{Classical $G$}

$G=SU(N)$ is  the simplest to analyze since possible values
of $n_s$ are either 1 or 0. With $H$ a proper subgroup of $SU(N)$, there has to be
at least one root that fails (\ref{H}), and  we can use the
Weyl transformation to bring it to the form
$$ n=(1,0,0,\dots ,0)$$
with $k_1^{(\Theta)}\notin {\mathbb Z}$. If all other
$k_s^{(\Theta)}$ are integral, $H\simeq SU(N-1)\times U(1)$, so
such a Wilson line does not contribute. Suppose that there is
exactly one more nonintegral $k_s^{(\Theta)}$. If $s> 2$,
the $U(1)$ persists and the saddle is irrelevant.
If $s=2$, and if $k_1^{(\Theta)}+ k_2^{(\Theta)}\notin {\mathbb Z}$,
the unbroken group $H$ has one more $U(1)$ factor. Finally,
if $k_1^{(\Theta)}+k_2^{(\Theta)}\in {\mathbb Z}$,
$H\simeq SU(2)\times U(1)\times SU(N-2)$, hence again
irrelevant. Proceeding similarly, it is easy to see that,
for $G=SU(N)$, an unbroken proper subgroup $H$ always
contains at least one $U(1)$ factor. Thus,
\bea
{\cal I}^{SU(N)}_{\rm bulk}(\beta z')\biggr\vert_{\beta\rightarrow 0}
= {\cal Z}^{SU(N)} (z') \ ,
\eea
as is consistent with Table \ref{numbers}.

The same table  also suggests, however, that {\it  for
no other simple Lie Group, such an equality will hold.}
For the  other classical groups, say $Sp(K)$ and $SO(N)$,
we find a large class of $H$-saddles, corresponding to
\bea
H\simeq Sp(m)\times Sp(K-m) \ , \qquad H\simeq SO(2m)\times SO(N-2m) \ .
\eea
The respective Wilson lines can be written compactly as
\bea
\Theta^{Sp(K)\rightarrow Sp(m)\times Sp(K-m)}  &= &\pi \sum_{s=1}^m{\mathbb H}_s \ , \cr\cr
\Theta^{SO(N)\rightarrow SO(2m)\times SO(N-2m)}&= &\pi \sum_{s=1}^m{\mathbb H}_s \ ,
\eea
which can be universally written as
\bea
k^{(\Theta)}_l =\frac12\delta_{lm}
\eea
with the canonical choice of simple roots,
\bea
\beta_1=e_1-e_2 \ , \quad \beta_2=e_2-e_3 \ ,\quad \cdots.
\eea
Absence of an unbroken $U(1)$ factor and $H\neq G$ further
demand $1\le m\le K-1$ for $Sp(K)$,
$4\le  2m \le  N-4$ for even $SO(N)$, and
$4\le  2m \le  N-1$ for odd $SO(N)$.

With these, the identity (\ref{correct}) simplifies to
\bea\label{Sp}
{\cal I}^{Sp(K)}_{\rm bulk}(\beta z')\biggr\vert_{\beta\rightarrow 0}
= \;{\cal Z}^{Sp(K)} (z')+\sum_{m=1}^{K-1} \frac14 \,{\cal Z}^{Sp(m)\times Sp(K-m)} (z')\ ,
\eea
and  similarly,
\bea\label{SOeven}
{\cal I}^{SO(N)}_{\rm bulk}(\beta z')\biggr\vert_{\beta\rightarrow 0}
= \;{\cal Z}^{SO(N)} (z')+\sum_{m=2}^{N/2-2} \frac{1}{8}\,{\cal Z}^{SO(2m)\times SO(N-2m)} (z')
\eea
for even $N$, and
\bea\label{SOodd}
{\cal I}^{SO(N)}_{\rm bulk}(\beta z')\biggr\vert_{\beta\rightarrow 0}
= \;{\cal Z}^{SO(N)} (z')+\sum_{m=2}^{(N-1)/2}\frac14\,{\cal Z}^{SO(2m)\times SO(N-2m)} (z')
\eea
for odd $N$.

These identities are checked affirmatively
by Table \ref{numbers}, and their analog for $\cN=8$ SYMQ.
For classical groups of general ranks, a conjectural
formula is available for the numerical limit of
${\cal Z}$ for $\cN=4, 8$  \cite{Pestun:2002rr}, while its 1d
counterpart ${\Omega}$'s has been also computed
\cite{Kac:1999av,Lee:2016dbm}. The comparison of these
two sets of numbers via the above identities is given
in the appendix. We also confirmed these for $\cN=16$
SYQM up to rank 3.

\subsection{Exceptional $G$}

$G_2$'s root system is generated by
the simple roots $\vec\beta_{1,2}$ with
$|\beta_2|^2=3|\beta_1|^2$ and $2\vec\beta_1\cdot
\vec\beta_2=-3|\beta_1|^2$.
The three short positive roots, $\{\beta_1,\beta_1+\beta_2,2\beta_1+\beta_2\}$,
and the three long positive roots, $\{\beta_2,3\beta_1+\beta_2,3\beta_1+2\beta_2\}$,
each span an $SU(3)$ root system.  $SU(3)$-saddles come from  Wilson lines
$\Theta$ with
\bea
k^{(\Theta)}=\left(\pm\frac13,0\right),
\eea
that leave the long roots unbroken,
while there are also $SU(2)\times SU(2)$-saddles at
\bea
k^{(\Theta)}=\left(0,\frac12\right), \quad k^{(\Theta)}=\left(\frac12,0\right),\quad k^{(\Theta)}=\left(\frac12,\frac12\right)
\eea
preserving $\{E_{\beta_1}, E_{3\beta_1+2\beta_2}\}$,
$\{E_{2\beta_1+\beta_2}, E_{\beta_2}\}$,
and $\{E_{\beta_1+\beta_2}, E_{3\beta_1+\beta_2}\}$, respectively.
Both classes of saddles will contribute, and (\ref{correct}) becomes
\bea
{\cal I}^{G_2}_{\rm bulk}(\beta z')\biggr\vert_{\beta\rightarrow 0}
=\; {\cal Z}^{G_2} (z')
+  \frac{1}{3}\,{\cal Z}^{SU(3)} (z')
+\frac12\, {\cal Z}^{SU(2)\times SU(2)} (z')\ ,
\eea
after we carefully keep track of the charge normalization
factors. This is, again, verified by Table \ref{numbers}.

For $F_4$, the root system is a
combination of $SO(9)$ roots and the 16 spinor weights thereof,
\bea
&&\pm \,2e_s\pm 2e_t \ ,\qquad s,t=1,2,3,4 \ ,\quad s\neq t \ ,\cr
&&\pm \,2e_s \ ,\qquad s=1,2,3,4 \ ,\cr
&&\pm \,e_1\pm e_2\pm e_3 \pm e_4 \ .
\eea
The following $2^3$ Wilson lines,
\bea
e^{i\Theta_{F_4}}=e^{i \sum_s \phi_s{\mathbb H}_s} \ ,\qquad \{e^{i\phi_s}\} =\{\pm 1, \pm 1,\pm 1, \pm 1\}
\eea
with an even number of $-1$'s leave the entire $F_4$ unbroken.
These saddles must be taken into account if we take the integration
range of each $\theta_s$ to be $[0,2\pi )$. As noted before, this
has something to do with the inherent ambiguity between
\bea
{\cal P}_{F_4}=\frac{1}{{\rm vol}(F_4)}\int_{F_4} e^{i\theta_a G_a}
\eea
and
\bea
{\cal P}_{F_4}'=\frac{1}{8{\rm vol}(F_4)}\int_{8F_4} e^{i\theta_a G_a}
\eea
where, in the latter, the Cartan torus is taken to have an
artificially enlarged volume $(2\pi)^4$. The latter choice
of the projector is implicitly used for the localization
computation of ${\cal I}^{F_4}_{\rm bulk}$, the factor 8
in the volume is correctly counteracted by this
eight-fold degeneracy of the $e^{i\Theta_{F_4}}$ saddles.

Thus, we have the associated 8-fold gauge-equivalence,
$({\mathbb Z}_2)^3$,  from the left multiplication by
$e^{i\Theta_{F_4}}$'s, up to which we  classify the $H$-saddles and
count $d_{F_4:H}$'s. The easiest
to spot are a triplet of $SO(9)$-saddles, modulo $({\mathbb Z}_2)^3$,
sitting at
\bea
 \{e^{i\phi_s}\} =\{ 1,  1,1, - 1\} \ ,\qquad \{i,i,i,i\} \ ,\qquad \{i,i,i,-i\}
\eea
respectively.
The first removes the $SO(9)$ spinor weights, while the
latter two removes $SO(8)$ vector weights and, respectively,
chiral or anti-chiral $SO(8)$ spinor weights. Thanks to
the $SO(8)$ triality, all of these preserve an $SO(9)$.
The Wilson lines
\bea
\{e^{i\phi_s}\} =\{ 1,  1,i, i\}
\eea
and the Weyl copies thereof, modulo $({\mathbb Z}_2)^3$,
produce $12$ $Sp(3)\times Sp(1)$ saddles.
Similarly,
\bea
\{e^{i\phi_s}\} =\{ \omega,  \omega ,\omega , \omega^3\}
\eea
with $\omega=e^{\pi i /4}$, produces  $24$  $SU(4)\times SU(2)$ saddles,
and
\bea
\{e^{i\phi_s}\} =\{ 1,\lambda , \lambda, \lambda^{2}\}
\eea
with $\lambda=e^{\pi i/3}$, produces $32$ $SU(3)\times SU(3)$ saddles.
These altogether imply, with (\ref{correct}),
\bea
{\cal I}^{F_4}_{\rm bulk}(\beta z')\biggr\vert_{\beta\rightarrow 0}&=&
{\cal Z}^{F_4} (z') + \frac{1}{2}\,{\cal Z}^{SO(9)} (z')+ \frac{1}{2}\,{\cal Z}^{Sp(3)\times Sp(1)} (z')\cr\cr
&&+\frac14\, {\cal Z}^{SU(4)\times SU(2)} (z')
+\frac13\, {\cal Z}^{SU(3)\times SU(3)} (z') \ ,
\eea
which is, again, easily confirmed by Table \ref{numbers}.

This leaves $G=E_{6,7,8}$. Since the above examples
illustrated and confirmed $H$-saddles and their consequences
amply, we will merely demonstrate existence of $H$-saddles for
these remaining cases. The root system of $E_8$ is
the combination of 112 roots and 128 chiral spinor weights of
$SO(16)$,
\bea
&&\pm\, 2e_s\pm2e_t \ ,\qquad  1\le s,t \le 8 \ , \quad s\neq t \ ,\cr
&&\pm\,  e_1\pm e_2\cdots \pm e_8 \ , \qquad \hbox{even number of } +\hbox{'s} \ .
\eea
A simple Wilson line,
\bea
\Theta =\pi\,{\mathbb H}_8
\eea
breaks the 128 entirely, while preserving $SO(16)$ subgroup,
so there are nontrivial $H\simeq SO(16)$ saddles.
For $E_7$, a useful representation of root systems is
\bea
&&\pm\, 2e_s\pm2e_t \ ,\qquad  1\le s,t \le 6 \ , \quad s\neq t \ ,\cr
&&\pm \, e_1\pm e_2\cdots\pm e_6 \pm \sqrt{2}\,e_7 \ , \qquad \hbox{even number of } +\hbox{'s for $e_{1,2,\dots,6}$} \ ,\cr
&& \pm\, 2\sqrt2 \,e_7\ ,
\eea
so the Wilson line
\bea
\Theta =\frac{\pi}{\sqrt 2}\;{\mathbb H}_7
\eea
preserves $H\simeq SO(12)\times SU(2)$. Finally, the $E_6$ root system is
\bea
&&\pm \,2e_s\pm2e_t \ ,\qquad  1\le s,t \le 5 \ , \quad s\neq t \ ,\cr
&&\pm \, e_1\pm e_2\cdots\pm e_5 \pm \sqrt{3}\,e_6 \ , \qquad \hbox{odd number of } +\hbox{'s} \ ,
\eea
and the Wilson line
\bea
\Theta =\frac{\pi}{4}\,\left({\mathbb H}_1+\cdots + {\mathbb H}_5 -\sqrt{3}\, {\mathbb H}_6 \right)
\eea
preserves $H\simeq SU(6)\times SU(2)$. 
So, there is at least one class of contributing $H$-saddles
for each of $E_{6,7,8}$ and,
\bea
{\cal I}^{E_{6,7,8}}_{\rm bulk}(\beta z')\biggr\vert_{\beta\rightarrow 0}
= \;{\cal Z}^{E_{6,7,8}} (z') +\cdots.
\eea
Again, one cannot resort to ${\cal Z}^{E_{6,7,8}} $ alone,
{\it a priori}, for the computation ${\cal I}^{E_{6,7,8}}_{\rm bulk}$.

\section{ADHM and 4d/5d Instanton Partition Functions}
\label{sec:ADHM}

Although we have so far  considered SYMQ with the adjoint
representation only, the same kind of problems in going
over from 1d to 0d can be expected generally. Perhaps
another most notable class is the ADHM-type that describes
D0 dynamics in D4 background, possibly with the additional
ingredients of Orientifold 4-planes or 8-planes and also
D8-branes. The 0d version enters instanton partition functions
for 4d supersymmetric Yang-Mills theories on Omega-deformed
${\mathbb R}^4$, such as the Nekrasov partition functions,
while the 1d version enters its 5d analog on
${\mathbb S}^1\times {\mathbb R}^4$.

It is clear that the phenomena of disappearing
residues will persist when we compare the 0d and the 1d
localization computations, since existence of $H$-saddles
originates in the vector multiplet;
the additional chiral or hypermultiplets can
only make such saddles more diverse than otherwise.
As with pure Yang-Mills cases, an $H$-saddle for twisted
partition functions of general gauged quantum mechanics means
a Wilson line $e^{i\Theta}$ at which the theory breaks up into
heavy and light parts and the resulting saddle point integral
survives the $\beta\rightarrow 0$ limit. One difference is that
this does not always require breaking of $G$ to a smaller
group; it may be that some of the matter fields can become
heavy instead. Thus, an $H$-saddle should generally refer to
contributing saddles at which the effective theory is
smaller than the one we started with.

This means that, sector by sector
labeled by the instanton number, the 0d and 1d partition
functions of the ADHM data cannot generally agree with each
other. On the other hand, we anticipate the instanton
partition functions are themselves very physical quantities,
easily expected to be continuous in the zero radius limit of
${\mathbb S}^1$. This is a qualitatively different issue
than the one we addressed so far: the new problem
emerges because ${\cal Z} ^{\rm ADHM}$'s
apparently have their own physical meanings, and the 4d/5d field
theory interpretations seem to require a continuity of some kind
with its 1d analog without having to add $H$-saddle contributions.

The resolution of this is already implicit in literatures.
It is well-known among practitioners that the ADHM
partition functions sometimes compute more than
what are needed for the instantons. The instanton
moduli space is equivalent to the Higgs phase, while
the ADHM themselves contain the Coulomb phase as well.
Since the latter is often factored out neatly, one
logical possibility is that the 1d/0d discontinuity
between  ${\cal I}_{\rm bulk}$ and
${\cal Z} $ resides entirely in the latter
Coulomb side and that way becomes irrelevant for the
field theory quantities. To see if this is actually
case, we will consider two distinct classes of ADHM
data. $U(k)$ ADHM data for $U(N)$ instantons
and $Sp(k)$ ADHM data for $SO(N)$ instantons.

\subsection{Missing Residues, Again}

Although $U(k)$ looks similar to $SU(k)$ superficially,
the two are very different. First of all, the distribution
of the adjoint poles in the 1d localization is such that
the $k$-fold degeneracy we have encountered in the adjoint-only
$SU(k)$ theory no longer appears. Indeed,
when we compute ADHM via 1d localization, it is clear that
there is no residue that would be dropped when we take the
strict 0d limit. As such, the 1d $U(k)$ ADHM partition
functions are continuously connected to 0d $U(k)$ ADHM
partition functions.

This can be seen more directly from $H$-saddle classification.
$U(k)$ ADHM field content differs from ${\cal N}=16$ $SU(N)$
theory in two respects: the overall $U(1)$ gauge group, always
unbroken under the Wilson line $\Theta$ and the $U(k)$ fundamental matter.
Since the latter couples to the $U(1)$ factors generically,
an unbroken $U(1)$ gaugino can easily find a Yukawa coupling
with another light fermion, apparently evading the condition
on allowed $H$-saddle. However, we need to recall that a
nontrivial Wilson line $\Theta$ always breaks $SU(k)$ part
of $U(k)$ with at least one factor of $U(1)$, so that the
unbroken group under $\Theta$ will break $U(k)$ to an $H$
with at least two factors of $U(1)$'s.
Furthermore, it is clear that light fermions
in the fundamental matter multiplet can couple to only one linear
combination of these two light gauginos. For example,
take $U(k)\rightarrow U(k')\times U(k-k')$
due to
$$
e^{i\Theta}={\rm diag}_{k\times k}(-1,-1,\dots, -1, 1,\dots,1) \ .
$$
The first $k'$ fermions in the fundamental become massive while
the other $k-k'$ remain massless. There are two light $U(1)$
gauginos, associated with each of the two blocks; While the
one associated with the $U(k-k')$ block can couple to
the light fundamental fermions, the other associated with
$U(k')$ can only couple to the heavy ones. The same is true of the
case $k=k'$; the single $U(1)$ gaugino couples to heavy fermions
only, and the saddle does not survive $\beta\rightarrow 0$ limit.
The same mechanism that killed potential $H$-saddles of pure
$SU(N)$ theory repeats itself for potential $H$-saddle for
$U(k)$ ADHM. The contributing saddle is possible only when
$H$ contains no more than one $U(1)$, or equivalently when
$H=U(k)=G$. Therefore, as we already noted based on the
localization comparison above, the matrix integral limit,
${\cal Z}^{U(k)}_{\rm ADHM}$, is continuously connected
to the respective 1d partition function $\Omega^{U(k)}_{\rm ADHM}$.

Let us now turn to $Sp(k)$ ADHM. The missing residue problem
appears already with $Sp(1)$, instead of starting with $Sp(2)$,
if fundamental matter is present. Proceeding with 1d
localization of $Sp(1)$ theory, we can expect to find
pairs of singularities, collectively denoted as $S^{(+)}$ and
$S^{(-)}$, where the adjoint charge becomes massless and ones,
say, $P$, involving fundamental becoming massless.
The pairs $S^{(\pm)}$ have positions mutually displaced by
the ${\mathbb Z}_2$ center; If we take $S^{(+)}$ to be the
ones that can be scaled to near the origin of $u$-space,
$S^{(-)}$ would be relatively displaced by $-1$ in $e^u$
coordinate.  These two would have different residues for
the ADHM case,
since the fundamental matter contributes differently.
The second type, $P$, would not be in pairs, as it is
entirely due to the fundamental matters. The passage
to 0d will then lose, $S^{(-)}$, while $S^{(+)}$ and $P$ can
be scaled to fit in $u'$ planes. Note that, even
though $S^{(-)}$ are related to $S^{(+)}$ by $Z_{Sp(1)}={\mathbb Z}_2$,
the unavoidable omission of $S^{(-)}$ cannot be cured by
an overall numerical factor of two on 0d side, unlike
the adjoint only $Sp(1)$ theory. With higher rank
$Sp(k)$, this problem will get only worse.

Such mismatches between the 1d and the 0d localizations
also manifest via contributing $H$-saddles in
$Sp(1)$ ADHM theory. What is the extra saddle? Consider
the Wilson line,
\bea e^{i\Theta}={\rm diag}_{2\times 2}(-1,-1) \eea
which is an element of $Z_{Sp(1)}={\mathbb Z}_2$.
The unbroken group $H$ equals $G=Sp(1)$
at this saddle, yet the fundamental multiplets become
massive  and the theory reduces to a
smaller one. This is clearly another form of the
$H$-saddle we discussed earlier in the section.

For $Sp(k)$ ADHM theories, therefore, the mismatch between
$\Omega^{Sp(k)}_{\rm ADHM}$ and ${\cal Z}^{Sp(k)}_{\rm ADHM}$
is unavoidable, already starting with rank 1.
Adding $H$-saddle contributions to the
latter will restore the limiting form of $\Omega^{Sp(k)}_{\rm ADHM}$,
but the problem at hand is that ${\cal Z}^{Sp(k)}_{\rm ADHM}$
might be by itself a physical quantity. The continuity between
5d and 4d instanton partition functions demands a different
kind of continuity between $\Omega^{Sp(k)}_{\rm ADHM}$ and
${\cal Z}^{Sp(k)}_{\rm ADHM}$, to which we turn next.

\subsection{Relating 4d and 5d}

As noted already, the crux of the matter lies in the fact
that ADHM partition function may compute more than what
are needed for the 4d/5d field theory instanton partition
functions. When D(-1) or D0 reside in D3/D4, they are the
field theory instantons. ADHM, however, knows about directions
transverse to D3/D4 as well,
in the form of the ``Coulomb branch." Since we are dealing
with low dimensional path integrals, the ``branches" are
not superselection sectors and have to be integrated
over as well. Therefore, the right thing to do is to factor
out the latter's contribution from the ADHM partition
functions, which should leave behind 4d/5d field theory quantities.

For $U(k)$ ADHM, it is known that the ``Coulomb" contributions
to the partition functions are absent. This can be understood
from the fact that the flavorless $U(k)$ theory comes
with a free $U(1)$. A free vector multiplet forces
${\cal I}$, $\Omega$, and ${\cal Z} $ to vanish
altogether. This implies the desired continuity,
\bea
\left[\sum q^k\Omega^{U(k)}_{\rm ADHM}\right]
\quad\rightarrow\qquad
\left[\sum q^k{\cal Z}^{U(k)}_{\rm ADHM}\right]\ ,
\eea
also reflected in the absence of $H$-saddle, or
equivalently in the absence of missing residue phenomena,
as discussed already.
On the other hand, for $Sp(k)$ and $O(m)$  ADHM data, the
continuity is lost. The Coulombic contributions in such
cases have been dealt with in the past in the context of
5d instantons \cite{Hwang:2014uwa,Hwang:2016gfw},
so what we need to do here is to check whether its
4d analog works as well. For example, is the limit
\bea
\left[\sum q^k\Omega^{Sp(k)}_{\rm ADHM}\right]
\left[\sum q^k\Omega^{Sp(k)}_{\rm Coulomb}\right]^{-1}\
\quad\rightarrow\quad
\left[\sum q^k{\cal Z}^{Sp(k)}_{\rm ADHM}\right]
\left[\sum q^k{\cal Z}^{Sp(k)}_{\rm Coulomb}\right]^{-1} \
\eea
continuous? Below, the simplest example of this for
4d $\cN=2^*$ and for 5d $\cN=1^*$ theories
is outlined for an illustration. A first-principle
computations of $\Omega^{Sp(k)}_{\rm ADHM}$'s were
already given in Ref.~\cite{Hwang:2016gfw}, while
their 4d counterpart, $\cZ^{Sp(k)}_{\rm ADHM}$'s,
are newly computed here for the above comparison.

The twisted partition function of the $Sp(k)$ ADHM
quantum mechanics is
\begin{align}
\label{eq:ind}
\Omega^{Sp(k)}_\text{ADHM} = \mathrm{Tr} \left[(-1)^\cF e^{-\beta
\{\cQ,\tilde\cQ\}} t^{2 (J_{1R}+J_{2R})} u^{2 J_{1L}} v^{2 J_{2L}}
\prod_{a = 1}^{N_f} w_a^{2 \Pi_a}\right],
\end{align}
where the supercharges are inherited from the 5d ones
as $\cQ \leftarrow  Q^{A = 1}_{\dot \alpha = \dot 1}$ and
$\tilde \cQ \leftarrow  Q^{A = 2}_{\dot \alpha = \dot 2}$,
while $J_{1R}$, $J_{1L}$, $J_{2R}$, and $J_{2L}$ are the
Cartan generators of $SU(2)_{1R}$, $SU(2)_{1L}$, $SU(2)_{2R},$ and
$SU(2)_{2L}$ $R$-symmetries, respectively. These
$SU(2)$'s sit inside the $SO(4)_1$ little group and
$SO(5)_2$ $R$-symmetry of the 5d theory in question.
$\Pi_a$'s are the Cartan generators of the flavor
symmetry, inherited from the 5d gauge symmetry $SO(2N_f)$.

With $N_f = 0$, this theory is a D0 quantum mechanics in
the presence of the $\text{O4}^{-}$-plane, and the partition
function here captures the ``Coulombic" part of the
ADHM data for $N_f\ge 1$. The generating function
for 1d ADHM partition functions has been proposed and
confirmed up to $k=3$ \cite{Hwang:2016gfw} as
\begin{align}
\label{eq:QM N=0}
\sum_{k = 0}^\infty q^k \Omega^{Sp(k)}_{\text{ADHM},SO(0)} &= \mathrm{PE} \left[\frac{1}{2} \frac{t^2 (v+v^{-1}-u-u^{-1}) (t+t^{-1})}{(1-t u) (1-t u^{-1}) (1+t v) (1+t v^{-1})} \frac{q}{1-q}\right].
\end{align}
where PE stands for the Plethystic exponential \cite{Feng:2007ur}.
Taking $\beta\rightarrow 0$ limit of this expression,
with the fugacities scaling as
\begin{align}
t = e^{-\beta \epsilon_+}, \quad u = e^{-\beta \epsilon_-}, \quad v = e^{-\beta m}, \quad w_i = e^{-\beta z_i}\ ,
\end{align}
we find
\begin{align}
\sum_{k = 0}^\infty q^k \Omega^{Sp(k)}_{\text{ADHM},SO(0)}\Biggr\vert_{\beta\rightarrow 0} = \;
\mathrm{exp} \left[\frac{1}{4} \frac{m^2-\epsilon_-^2}{\epsilon_+^2-\epsilon_-^2} \sum_{n = 1}^\infty \frac{1}{n}\frac{q^n}{1-q^n}\right]\ .
\end{align}
On the other hand, our ADHM matrix integrals suggest
 \begin{align}
\label{eq:mat N=0}
\sum_{k = 0}^\infty q^k \mathcal Z^{Sp(k)}_{\text{ADHM},SO(0)} &= \mathrm{exp} \left[\frac{1}{8} \frac{m^2-\epsilon_-^2}{\epsilon_+^2-\epsilon_-^2} \sum_{n = 1}^\infty \frac{1}{n}\frac{q^n}{1-q^n}\right],
\end{align}
instead. It is clear that the $\beta\rightarrow 0$
limit of \eqref{eq:QM N=0} does not match \eqref{eq:mat N=0},
as anticipated from  our $H$-saddle story.

Similarly, the 1d twisted partition functions for $N_f=1$ are
also proposed and checked up to $k=3$ by the authors of
Ref.~\cite{Hwang:2016gfw} as
\begin{align}
\label{eq:QM N=2}
\sum_{k = 0}^\infty q^k \Omega^{Sp(k)}_{\text{ADHM},SO(2)} &= \mathrm{PE} \left[\frac{1}{2} \frac{t^2 (v+v^{-1}-u-u^{-1}) (2 v+2 v^{-1}+3 t+3 t^{-1})}{(1-t u) (1-t u^{-1}) (1+t v) (1+t v^{-1})} \frac{q}{1-q}\right]
\end{align}
with $\beta\rightarrow 0$ limit,
\begin{align}
\sum_{k = 0}^\infty q^k \Omega^{Sp(k)}_{\text{ADHM},SO(2)}\Biggr\vert_{\beta\rightarrow 0} = \;
\mathrm{exp} \left[\frac{5}{4} \frac{m^2-\epsilon_-^2}{\epsilon_+^2-\epsilon_-^2} \sum_{n = 1}^\infty \frac{1}{n}\frac{q^n}{1-q^n}\right]\ .
\end{align}
We find that the ADHM matrix integrals are consistent with
\begin{align}
\label{eq:mat N=2}
\sum_{k = 0}^\infty q^k \mathcal Z^{Sp(k)}_{\text{ADHM},SO(2)} &= \mathrm{exp} \left[\frac{9}{8} \frac{m^2-\epsilon_-^2}{\epsilon_+^2-\epsilon_-^2} \sum_{n = 1}^\infty \frac{1}{n}\frac{q^n}{1-q^n}\right].
\end{align}
so, again, there is a discrepancy between the two.

If this discrepancy lies entirely in the ``Coulombic"
part of the ADHM data, the continuity would be restored
by taking a ratio between $N_f=1$ and $N_f=0$ generating
functions. Indeed, from the comparison of the above four
generating functions, it is pretty clear that this is the
case with
\bea
\left[\sum q^k{\cal Z}^{Sp(k)}_{{\rm ADHM},SO(2)}\right]
\left[\sum q^k{\cal Z}^{Sp(k)}_{{\rm ADHM},SO(0)}\right]^{-1} =\; \mathrm{exp} \left[\frac{m^2-\epsilon_-^2}{\epsilon_+^2-\epsilon_-^2} \sum_{n = 1}^\infty \frac{1}{n} \frac{q^n}{1-q^n}\right]\ .
\eea
We have performed a similar check for $SO(N\le 5)$
instantons, again up to $k=3$, confirming the proposed
continuity convincingly.

We need to emphasize one important normalization
issue that enters the comparison. Recall that for
pure Yang-Mills cases, the matrix side must be
multiplied by an overall factor $|{\rm det}(Q^G)|$,
which corrects the problem that the 0d localization
misses the pole locations shifted by $Z_G$. On the
other hand, with ADHM data, there is no such obvious
numerical factor. Instead, as we saw in the $Sp(1)$
case, a $Z_G$-shifted missing residue is inequivalent
to the one at origin, and should be considered a
nontrivial $H$-saddle. This example suggests that
the multiplicative factor for ${\cal Z}^{G}_{\rm ADHM}$
should be the one associated with the fundamental
charges, say, $|{\rm det}(Q^G_F)|=1$ for classical $G$'s.
It is with this normalization choice
that the proposed continuity between 5d and 4d works.

An interesting exercise would be to study how
these features manifest in the instanton partition
functions for general 4d $\cN=2$ theories and for
general 5d $\cN=1$ theories \cite{Nekrasov:2002qd,Nekrasov:2003rj,Nekrasov:2004vw}.
Computations for these objects have been carried out
extensively in the past,
yet some ambiguity seems to persist for the symplectic
and the orthogonal cases. A more thorough investigation
of 4d/5d instanton partition functions in view of
our new findings will appear elsewhere.

\vskip 0.5cm

\section*{Acknowledgement}
We would like to thank Richard Eager for bringing our
attention to Ref.~\cite{Krauth:2000bv}, and also Joonho Kim
and Jaewon Song for useful discussions. P.Y. is grateful
to Seung-Joo Lee, previous collaborations with whom
motivated the main question of this note.

\vskip 1cm
\appendix
\section{$\Omega^G$ and $\cZ^G$ for High Rank Classical Groups}

A general and algebraic expression for $\Omega^G$ with $\cN= 4,8$
and any simple Lie group $G$ is known in terms of the Weyl group
$W_G$ as follows
\cite{Lee:2016dbm},
\bea\label{defect}
\Omega^{G}_{\cN=4}&=& \frac{1}{|W_{G}|}\sum'_{w}\frac{1}{{\rm det}
\left({\bf y}^{-1}-{\bf y}\cdot w\right)}\ ,
\cr\cr
\Omega^{G}_{\cN=8}&=& \frac{1}{|W_{G}|}\sum'_{w}\frac{1}{{\rm det}
\left({\bf y}^{-1}-{\bf y}\cdot w\right)}  \cdot
\frac{{\rm det}\left(x^{F/2}{\bf y}^{-1}-x^{-F/2}{\bf y}
\cdot w\right)}{{\rm det}\left( x^{F/2}
- x^{-F/2}\cdot w\right)}\ ,
\eea
with the common unrefined limit \cite{Yi:1997eg,Green:1997tn,Kac:1999av}
\bea\label{unrefined}
\Omega^{G}&=& \frac{1}{|W_{G}|}\sum'_{w}\frac{1}{{\rm det}
\left(1- w\right)}\ ,
\eea
where the sum is restricted to the so-called elliptic Weyl elements,
defined by ${\rm det}(1-w)\neq 0$.

The formula (\ref{unrefined}) has been further evaluated
for classical groups as \cite{Pestun:2002rr},
\bea\label{KS}
&&\Omega^{Sp(r)}=\Omega^{SO(2r+1)}= \frac{1}{2^{2r}r!}\prod_{j=0}^{r-1}(4j+1)\ , \cr\cr
&&\Omega^{SO(2r)}= \frac{1}{2^{r-1} r!} \, 2^{-r-1} \left(\prod_{j=0}^{r-1}(4j+1)+\prod_{j=0}^{r-1}(4j-1)\right).
\eea
On the other hand, the same reference offered conjectural
formulae for the matrix integral counterpart.
\bea\label{Pestun}
&&\cZ^{Sp(r)}=\frac{1}{2^{3r-1}r!}\prod_{j=1}^{r-1}(8j+1) \ ,  \cr\cr
&&\cZ^{SO(2r+1)}= \frac{1}{2^r r!} \sum_{l = 1}^r 2^{r+1-3 l} F_r^l b_l \ , \cr\cr
&&\cZ^{SO(2r)}= \frac{2}{2^{r-1} r!} \sum_{l/2 = 1}^{[r/2]} 2^{r+1-3 l} F_r^l b_l\ .
\eea
$F_r^l$ is the absolute value of the Stirling number of the
first kind, or equivalently
\begin{align}
\sum_{l=1}^r F_r^l s^l = s(s+1)(s+2)\cdots(s+r-1)\ ,
\end{align}
while $b_l$ is a sequence of integers defined by
\begin{align}
b_{2 k} = b_{2 k-1} = (-1)^{k+1} 2^{k-1} \beta_k
\end{align}
where $\beta_k$  is from the $k$-th expansion coefficient of $\sqrt{\cos x}$:
\begin{align}\label{beta}
\sqrt{\cos x} = 1-\sum_{k = 0}^\infty \frac{\beta_k  x^{2 k}}{2^k (2 k)!} \ .
\end{align}
For $SO$'s, by the way, our expressions actually differ
a little from those in Ref.~\cite{Pestun:2002rr}, in part
due to various typos in the latter. The above expressions
(\ref{KS}) and (\ref{Pestun}) are consistent with low rank
numbers in Table 1.

Now we are ready to check the identities (\ref{Sp}-\ref{SOodd})
against for classical groups of general ranks.
For $SO$ groups, we have confirmed
(\ref{SOeven}-\ref{SOodd}) numerically against (\ref{KS})
and (\ref{Pestun}), up to the rank 100, going well beyond Table 1.
For the symplectic cases, a general and elementary
proof follows once we rewrite (\ref{Sp}) as
\bea \label{Sp'}
\Omega^{Sp(K)}
= \frac14 \sum_{r=0}^{K} \,{\cal Z}^{Sp(r)}\cdot {\cal Z}^{Sp(K-r)}\ ,
\eea
with $\cZ^{Sp(0)}\equiv 2$ understood as a natural extrapolation
from (\ref{Pestun}). Numerical limits of
both $\cZ$'s and $\Omega$'s are  proportional to binomial
coefficients as
\bea
&&\frac{\cZ^{Sp(r)}}{2}
= \frac{\Gamma(\frac18+r)}{r!\Gamma(\frac18)}
= (-1)^r\left(\begin{array}{r} -\frac18\\ r\end{array}\right)
\ , \cr\cr
&&\Omega^{Sp(r)}
= \frac{\Gamma(\frac14+r)}{r!\Gamma(\frac14)}
= (-1)^r\left(\begin{array}{r}-\frac14\\ r\end{array}\right)\ ,
\eea
and, as such, the identity (\ref{Sp'}) follows
immediately from the trivial equality,
$$(1+s)^{-1/8} \cdot (1+s)^{-1/8} = (1+s)^{-1/4} \ , $$
i.e., from the binomial expansions thereof.

Recall that we have derived the identities (\ref{Sp}-\ref{SOodd})
rigorously  from the path integral while the
general formulae (\ref{defect}), also confirmed by explicit
localization computations  up to rank 7, have strong physical
motivations \cite{Yi:1997eg,Kac:1999av,Lee:2016dbm}.
Given these, it is perhaps more sensible to view these checks
as a confirmation of the numerics  in (\ref{Pestun}).

\end{document}